\documentclass[11pt]{article}
\usepackage{amssymb}
\def\Journal#1#2#3#4{{#1} {\bf #2}, #3 (#4)}
\def\PLB{{\em Phys. Lett.}  B}

\def\JPA{{\em J. Phys.} A}
\def\JMP{\em J. Math. Phys.}
\def\CMP{\em Commun. Math. Phys.}
\def\LMP{\em Lett. Math. Phys.}
\def\IJMPAp{\em Int. J. Mod. Phys. A (Proc. Suppl.)}
\def\CRASP{\em C. R. Acad. Sci. Paris}

\def\ra{\rightarrow}
\def\be{\begin{equation}}
\def\ee{\end{equation}}
\def\bea{\begin{eqnarray}}
\def\eea{\end{eqnarray}}
\def\bee{\begin{eqnarray*}}
\def\eee{\end{eqnarray*}}
\def\O{\Omega}
\def\A{{\cal A}}
\def\M{{\cal M}}
\def\d{\mbox{d}}

\def\pa{\partial}

\newcommand{\idty}{{\leavevmode{\rm 1\mkern -5.4mu I}}}
\begin{document}

\begin{tabbing}
\hspace*{11cm} \= GOET-TP 99/97  \\
               \> November 1997               \\
               \> 
\end{tabbing}

\centerline{\Large \bf SOME ASPECTS OF NONCOMMUTATIVE }
\vskip.2cm
\centerline{\Large \bf GEOMETRY AND PHYSICS}
\vskip1cm
\begin{center}
      {\bf A. DIMAKIS$^1$} and {\bf F. M\"ULLER-HOISSEN$^{2,3}$} 
       \vskip.3cm 
      \begin{minipage}{15cm}
      $^1$Department of Mathematics, Univ. of the Aegean,
      GR-83200 Karlovasi, Samos, Greece
      \vskip.1cm 
      \noindent
      $^2$Institut f\"ur Theoretische Physik,
      Bunsenstr. 9, D-37073 G\"ottingen, Germany 
      \vskip.1cm
      \noindent
      $^3$Max-Planck-Institut f\"ur Str\"omungsforschung,
      Bunsenstr. 10, D-37073 G\"ottingen     
      \end{minipage}
\end{center} 
\vskip1cm

\abstract{An introduction is given to some selected aspects of noncommutative geometry.
Simple examples in this context are provided by finite sets and lattices. As an application,
it is explained how the nonlinear Toda lattice and a discrete time version of it can be understood 
as generalized $\sigma$-models based on noncommutative geometries. In particular, 
in this way one achieves a simple understanding of the complete integrability of the Toda 
lattice. Furthermore, generalized metric structures on finite sets and lattices are briefly 
discussed.}

\section{Introduction}
Noncommutative (differential) geometry extends notions of classical differential geometry
from differentiable manifolds to discrete spaces, like finite sets and fractals, and even
`noncommutative spaces' (`quantum spaces') which are given by noncommutative associative 
algebras (over $\mathbb{R}$ or $\mathbb{C}$). Such an algebra $\A$ replaces the commutative 
algebra of  $C^\infty$-functions on a smooth manifold $\M$. 
\vskip.2cm

A basic concept of classical differential geometry is the notion of a vector 
field. The latter is a derivation $C^\infty(\M) \ra C^\infty(\M)$. One may think of generalizing
the notion of a vector field as a derivation $\A \ra \A$ to associative algebras.\cite{Dubo88} 
But there are interesting algebras, like the (commutative) algebra of functions on a finite set, 
on which there is no derivation at all except the trivial one (which maps all functions to zero).
It seems that there is no adequate general definition of a vector field. 
\vskip.2cm

Perhaps of even more importance than vector fields are differential forms in classical
differential geometry and theoretical physics. They extend the algebra $\O^0(\M) := C^\infty(\M)$
to a graded associative algebra $\O(\M) = \bigoplus_{r \geq 0} \O^r(\M)$, the algebra of 
differential forms. A central part of its structure is encoded in the action of the
exterior derivative $\d \, : \, \O^r(\M) \ra \O^{r+1}(\M)$ which is a (graded) derivation. 
Given any associative algebra $\A$, one can always associate with it a differential algebra
$\O(\A)$ which should be regarded as an analogue of the algebra of classical differential 
forms on a manifold. Further geometric structures, like connections and tensors, can then be 
built on such a differential algebra in very much the same way as in classical differential
geometry.
\vskip.2cm

There is no universal way of associating a unique differential algebra with a given associative
algebra $\A$. In general, there is no distinguished differential algebra and one has to understand 
what the significance is of the different choices. Even in the case of the algebra of smooth functions 
on a manifold, there is actually no longer a good argument to single out the ordinary calculus of 
differential forms. The latter is distinguished, however,  via our classical conception how to measure 
volumes. The exploration of other differential calculi opened a door to a  new world of geometry and 
applications in physics, as well as relations with other fields of mathematics. Some aspects have been
briefly reviewed in a recent paper \cite{MH97} to which we refer for further 
information.\footnote{At present, this can be accessed online via 
{\em http://kaluza.physik.uni-konstanz.de/2MS}. } 
\vskip.2cm

The formalism of noncommutative geometry is an extremely radical abstraction 
of ordinary differential geometry. It includes the latter as a special case, 
but allows a huge variety of different structures. In particular, it is possible to `deform' ordinary 
differential geometry and models built on it while keeping basic concepts and recipes on 
which the models are based. 
\vskip.2cm

In the following we review some aspects of noncommutative geometry concentrating
on a few easily accessible examples. Section 2 collects basic definitions
of differential calculus on associative algebras. Section 3 recalls some facts
about differential calculus on finite sets and a correspondence between first order
differential calculi and digraphs.\cite{DMH94_ddm} 
Also a relation with an approach of Alain Connes \cite{Conn94} to noncommutative 
geometry is explained. 
Section 4 treats differential calculus
on linear and hypercubic lattices. The corresponding differential calculi may
be regarded as deformations of the ordinary differential calculus on ${\mathbb{R}}^n$.
A more general class of such differential calculi is briefly discussed in section
5. The `lattice differential calculus' underlies an important example of a
discrete `generalized $\sigma$-model'. As explained in section 6, such models 
generalize a class of completely integrable two-dimensional classical 
$\sigma$-models \cite{BIZZ79} by replacing the ordinary differential calculus 
by a noncommutative one.\cite{DMH96_intd,DMH97_intm}
They involve a generalized Hodge $\star$-operator. In classical 
(Riemannian) differential geometry, the Hodge operator is obtained from a metric
which defines the distance between two points of the manifold. Distance
functions on a rather general class of algebras have been introduced by
Connes.\cite{Conn94} In particular, his definition applies to discrete sets. But the
relation with a Hodge operator has still to be worked out. Section 7 is devoted to
a discussion of some metric aspects. Section 8 contains some final remarks.

\section{Differential calculi on associative algebras}
Let $\A$ be an associative algebra over $\mathbb{R}$ or $\mathbb{C}$ with unit $\idty$. 
A {\em differential calculus} on $\A$ is a $\mathbb{Z}$-graded 
associative algebra (over $\mathbb{R}$, respectively $\mathbb{C}$)\footnote{Though in many
interesting cases one has $\O^r(\A) = \lbrace 0 \rbrace$ when $r$ is larger
than some $r_0 \geq 0$, one encounters examples where $\O (\A)$ is
actually an infinite sum. $\O (\A)$ is then the space of all finite sums of arbitrary order.}
\be
           \O (\A) = \bigoplus_{r \geq 0} \O^r (\A)
\ee 
where the spaces $\O^r (\A)$ are $\A$-bimodules\footnote{The elements of $\O^r (\A)$,
called $r$-{\em forms}, can be multiplied from left and right by elements of 
$\A$.} 
and $\O^0(\A) = \A$. There is a ($\mathbb{R}$- respectively $\mathbb{C}$-) linear map  
\be
     \d \; : \quad \O^r (\A) \rightarrow \O^{r+1}(\A)
\ee
with the following properties,
\bea
       \d^2 &=& 0 \\
       \d (w \, w') &=& (\d w) \, w' + (-1)^r \, w \, \d w'
                           \label{Leibniz}
\eea
where $w \in \O^r(\A)$ and $w' \in \O (\A)$. The last relation is
known as the (generalized) {\em Leibniz rule}. One also requires 
$\idty \, w = w \, \idty = w$ for all elements $w \in \O (\A)$. The 
identity $\idty \idty = \idty$ then implies
\be
            \d \idty = 0   \; .
\ee 
We assume that $\d$ generates the spaces $\O^r(\A)$ 
for $r>0$ in the sense that $\O^r(\A) = \A \, \d \O^{r-1}(\A) \, \A$.
Using the Leibniz rule, every element of $\O^r(\A)$ can be
written as a linear combination of monomials 
$a_0 \, \d a_1 \cdots \d a_r$. The action of $\d$ is then
determined by
\be
 \d ( a_0 \, \d a_1 \cdots \d a_r ) = \d a_0 \, \d a_1 \cdots \d a_r
  \; .
\ee 
So far nothing has been said about commutation relations for elements
of $\A$ and differentials. Indeed, in the largest differential calculus, the
{\em universal differential envelope} $( \tilde{\O}(\A) , \tilde{\d} )$ 
of $\A$, there are no such relations.
Smaller differential calculi are obtained by specifying corresponding
commutation relations (which have to be consistent with the existing
relations in the differential algebra, of course). The smallest differential
calculus is $\O(\A) = \A$ where $\d$ maps all elements of $\A$ to
zero.

\section{Differential calculi on a finite set}
Let $\cal M$ be a finite set and $\A$ the algebra of all $\mathbb{C}$-valued functions
on it. $\A$ is generated by $\lbrace e_i \rbrace$ where $e_i(j) = \delta_{ij}$
for $i,j \in \cal M$. These functions satisfy the two identities
\be        \label{fin_set_rels}
     e_i \, e_j = \delta_{ij} \,  e_j \, , \qquad 
  \sum_i e_i = \idty  \; .
\ee
As a consequence of the identities (\ref{fin_set_rels}) and the Leibniz rule, the 
differentials $\d e_i$ of a differential calculus on $\A$ are subject to the 
following relations,
\be
      \d e_i \; e_j  = - e_i \, \d e_j + \delta_{ij} \, \d e_j  \, , \qquad 
    \sum_i \d e_i = 0    \; .
\ee
Without additional constraints, we are dealing with the universal differential 
calculus $( \tilde{\O}(\A) , \tilde{\d} )$. Introducing the 1-forms
\be
       e_{ij} = e_i \, \tilde{\d} e_j  \qquad  (i \neq j)
\ee
one finds that they form a basis over $\mathbb{C}$ of the space $\tilde{\O}^1$ of 
universal 1-forms. Moreover, all first order differential calculi are obtained 
from the universal one by setting some of the $e_{ij}$ to zero.

\vskip.3cm
Let us associate with each nonvanishing $e_{ij}$ of some differential calculus 
$(\O,\d)$ an arrow from the point $i$ to the point $j$ :
\be
    e_{ij} \neq 0 \quad \Longleftrightarrow \quad i \, \bullet \, \longrightarrow 
    \, \bullet \, j     \; .
\ee 
The universal (first order) differential calculus then corresponds to the complete 
digraph where all vertices are connected with each other by a pair of antiparallel
arrows. Other first order differential calculi are obtained by deleting some
of the arrows.
The choice of a (first order) differential calculus on a finite set therefore means assigning
a connection structure to it. The latter is mirrored in the following formula for the 
differential of a function $f \in \A$,\footnote{More precisely, the summation runs over all $i,j$ with $i \neq j$.
Note that $e_{ii}$ has not been defined. We may, however, set $e_{ii} := 0$.} 
\be                              \label{df_e_ij}
         \d f = \sum_{i,j}  \lbrack f(j) - f(i) \rbrack \,  e_{ij}      \; .
\ee

Returning to the universal calculus, concatenation of the 1-forms $e_{ij}$ leads to the 
{\em basic $(r-1)$-forms}
\be                                        \label{e_...}
        e_{i_1 \ldots i_r} := e_{i_1 i_2} \, e_{i_2 i_3} \cdots e_{i_{r-1} i_r}  
        \qquad  ( r > 1 )  \; .
\ee 
They constitute a basis of $\tilde{\O}^{r-1}$ over $\mathbb{C}$ and
satisfy the simple relations
\be                          \label{e_e_delta_e}
       e_{i_1 \ldots i_r} \, e_{j_1 \ldots j_s} 
   = \delta_{i_r j_1} \, e_{i_1 \ldots i_{r-1} j_1 \ldots j_s}  \; .
\ee
  Furthermore, we have
\bea
      \tilde{\d} e_i &=& \sum_j ( e_{ji} - e_{ij} )   \\
      \tilde{\d} e_{ij} &=& \sum_k ( e_{kij} - e_{ikj} + e_{ijk} )   \\
      \tilde{\d} e_{ijk} &=& \sum_l ( e_{lijk} - e_{iljk} + e_{ijlk} - e_{ijkl} ) \\
                        & \vdots &   \nonumber  
\eea
The first equation is a special case of (\ref{df_e_ij}). In a `reduced'
differential calculus $(\O, \d)$ where not all of the $e_{ij}$ are present, the 
possibilities to build (nonvanishing) higher forms $e_{i_1 \ldots i_r}$ are 
restricted and the above formulas for $\tilde{\d} e_{i_1 \ldots e_r}$ impose further 
constraints on them. 

\vskip.3cm
\noindent
{\em Example.} The graph drawn in Fig. 1 determines a first order differential calculus
with nonvanishing basic 1-forms $e_{12}, e_{23}, e_{14}, e_{43}$. Concatenation
only leads to $e_{123}$ and $e_{143}$ as possible nonvanishing basic 2-forms.
There are no nonvanishing $r$-forms with $r>2$. 

\begin{minipage}{7cm}
\begin{center}
\unitlength=1.5cm     
\begin{picture}(4.,1.8)(-0.5,-0.4)
\thicklines
\put(0.,0.) {\circle*{0.1}}
\put(1.,0.) {\circle*{0.1}}
\put(0.,1.) {\circle*{0.1}}
\put(1.,1.) {\circle*{0.1}}
\put(-0.2,-0.05) {$1$}
\put(1.15,-0.05) {$2$}
\put(1.15,0.95) {$3$}
\put(-.2,0.95) {$4$}
\put(0.,0.) {\vector(1,0){0.9}}
\put(1.,0.) {\vector(0,1){0.9}}
\put(0.,0.) {\vector(0,1){0.9}}
\put(0.,1.) {\vector(1,0){0.9}}
\end{picture}
\end{center}
\end{minipage}
\hfill
\begin{minipage}{4cm}
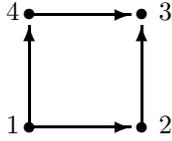
\centerline{\bf Fig. 1}
\small
\noindent
The digraph associated with a special differential calculus on a set 
of four elements.
\end{minipage}

\noindent
The graph is obtained from the complete 
digraph by deletion of some arrows. In particular, an arrow from point $1$ to point $3$
is missing which corresponds to setting $e_{13}$ to zero in the universal differential calculus. 
This leads to $ 0 = \d e_{13} =  - e_{123} - e_{143}$. 
                                                                       \hfill $\blacksquare$

\vskip.3cm

A discrete set together with a differential calculus on it is called
a {\em discrete differential manifold}. \cite{DMHV95}

\subsection{Representations of first order differential calculi on finite sets}
As explained above, first order differential calculi on a set of $N$ elements are 
in bijective correspondence with digraphs with $N$ vertices and at most 
a pair of antiparallel arrows between any two vertices. On the other hand,
in graph theory such a digraph is characterized by its {\em adjacency matrix} which 
is an $N \times N$-matrix ${\cal D}$ such that $ {\cal D}_{ij} = 1$ if there is an 
arrow from $i$ to $j$ and $ {\cal D}_{ij} = 0$ otherwise. 
One should then expect that the (first order) differential calulus determined by a 
digraph can be expressed in terms of $\cal D$. The simplest way to build a derivation 
$\d \, : \, \A \rightarrow \O^1(\A)$ is as a commutator,
\be
        \d f := \lbrack {\cal D} , f \rbrack     \label{df_D}
\ee 
which presumes, however, that the elements of $\A$ can be represented as 
$N \times N$-matrices. But this is naturally achieved via
\bea                     \label{f_repr}
      f   \quad \mapsto \quad \left( \begin{array}{ccc} f(1) &  & 0 \\
                                              & \ddots   &  \\
                                           0 &               & f(N)  
                                              \end{array} \right )   \; .
\eea 
Comparison of (\ref{df_D}) with our formula (\ref{df_e_ij}) shows that
the basic 1-form $e_{ij}$ is represented as the $N \times N$-matrix $E_{ij}$ with 
a $1$ in the $i$th row and $j$th column and zeros elsewhere.
The adjacency matrix $\cal D$ represents $\sum_{i,j} e_{ij}$.
\vskip.3cm

Proceeding beyond 1-forms, the above representation will not respect
the ${\mathbb{Z}}_2$-grading of a differential algebra $\O(\A)$. One may
consider instead a `doubled' representation \cite{DMH94_fin}
\bea
 e_i \, \mapsto \, \left( \begin{array}{cc} E_{ii} & 0 \\ 0 & E_{ii} \end{array}  
        \right) \, , \qquad
 e_{ij} \, \mapsto \, \left( \begin{array}{cc} 0 & E_{ij}^\dagger \\ 
                                          E_{ij} & 0
        \end{array} \right)   \; .      \label{doubled_repr}
\eea
The grading can be expressed in terms of a grading operator which in 
our case is given by
\bea
         \gamma := \left( \begin{array}{cc}  {\bf 1} & 0 \\ 0 & - {\bf 1}
                                     \end{array} \right)  \; .
\eea
It is selfadjoint and satisfies
\be
      \gamma^2 = {\bf 1} \, , \qquad
      \gamma \, \hat{\cal D} = - \hat{\cal D} \, \gamma \, \qquad 
      \gamma \, \hat{f} = \hat{f} \, \gamma 
\ee 
with 
\bea        \hat{\cal D} :=  \left( \begin{array}{cc} 0 & {\cal D}^\dagger \\
                       {\cal D} & 0 \end{array} \right)   \, \qquad
                \hat{f} :=  \left( \begin{array}{cc} f & 0 \\ 0 & f
                                         \end{array} \right)
                     \label{doubled_D,f}
\eea
where $f$ has to be represented as in (\ref{f_repr}). In this way we
do {\em not} in general obtain a representation of the first order 
differential calculus which we started with, however, but a representation
of the corresponding `symmetric' differential calculus where with
$e_{ij} \neq 0$ there is also $e_{ji} \neq 0$ (so that the associated
digraph is symmetric). 
\vskip.3cm

With the above representations of (first order) differential calculi we have
established contact with Alain Connes' formalism \cite{Conn95} of noncommutative 
geometry. But  in the present context $\cal D$ is {\em not}, in general, a {\em selfadjoint} 
operator (on the Hilbert space ${\mathbb{C}}^N$). The `doubling' in (\ref{doubled_repr}) 
leads to a selfadjoint operator on the Hilbert space ${\cal H} = {\mathbb{C}}^{2N}$, 
however. 
$(\A, {\cal H}, \hat{\cal D})$ is an example of an {\em even spectral triple},
a central structure in Connes' approach to noncommutative geometry.\cite{Conn95,Conn96}
A {\em spectral triple} $(\A, {\cal H}, \hat{\cal D})$ consists of an involutive 
algebra $\A$ of operators on a Hilbert space $\cal H$ together with a selfadjoint 
operator $\hat{\cal D}$ satisfying some technical conditions. It is called 
{\em even} when there is a grading operator $\gamma$, as in our example.

\section{Lattice differential calculus}
Let $\M = \mathbb{Z}$. For $i,j \in \M$ we define a differential calculus
by $ e_{ij} \neq 0 \;  \Longleftrightarrow \;  j = i + 1$ following
the rules described for finite sets in the previous section. 
This corresponds to the oriented linear lattice graph drawn in Fig. 2.

\begin{minipage}{4cm}
\unitlength=1.cm
\begin{picture}(6,1)(-0.5,0)
\thicklines
%\linethickness{0.3mm}
%
\put(0.3,0){$\ldots$}
\multiput(1,0)(1,0){6}{\circle*{0.15}}
\multiput(1,0)(1,0){5}{\vector(1,0){0.9}}
\put(6.3,0){$\ldots$}
\end{picture}
\end{minipage}
\hfill
\centerline{
\begin{minipage}[t]{2.6cm}
\centerline{\bf Fig. 2}
\vskip.1cm \noindent
\small
An oriented linear lattice graph. 
\end{minipage}    }
 
\vskip.2cm
\noindent
In this example we are dealing with an {\em infinite}
set and thus infinite sums in some formulas which 
would actually require a bit more care. 
Introducing the lattice coordinate function
\be
          x := \ell \, \sum_j j \, e_j
\ee
with a real constant $\ell >0$, one obtains 
\bea
  \d x = \ell  \sum_i i \, \d e_i = \ell  \sum_{i,j} i \, (e_{ji} - e_{ij}) 
         = \ell  \sum_i i \, (e_{i-1,i} - e_{i,i+1}) = \ell  \sum_i e_{i,i+1}
        \qquad
\eea   
and 
\bea
 \lbrack \d x , x \rbrack = \ell^2 \, \sum_{i,j} j \, \lbrack e_{i,i+1} , e_j \rbrack   
     = \ell^2 \,  \sum_i e_{i,i+1} = \ell \, \d x   
\eea
using (\ref{e_e_delta_e}). Hence
\be
           \lbrack \d x , x \rbrack = \ell \, \d x   \; .     \label{dx_x_latt}
\ee
In the limit $\ell \to 0$ the lattice coordinate function $x$ naively approximates
the corresponding coordinate function on the real line  $\mathbb{R}$. From our
last equation we then recover the familiar commutativity of ordinary differentials
and functions. The above commutation relation makes also sense, however,
on  $\mathbb{R}$ when $\ell > 0$. We then have a {\em deformation} of the ordinary
calculus of differential forms on  $\mathbb{R}$ with deformation parameter $\ell$.
In the following we collect some properties of this deformed differential calculus.
Written in the form
\be
       \d x \, x = (x + \ell) \, \d x  \, ,     
\ee
the above commutation relation extends to the algebra $\A$ of all functions on 
$\mathbb{R}$ as
\be
    \d x \, f(x) = f(x + \ell) \, \d x  \; .       \label{dx_f_1dlatt}
\ee 
 Furthermore,
\bea
  \d f &=:& (\partial_{+x} f) \, \d x 
        =  {1 \over \ell} \, (\partial_{+x} f) \, \lbrack \d x , x \rbrack      
        =  {1 \over \ell} \, \lbrack (\partial_{+x} f) \,  \d x , x \rbrack  \nonumber \\
     &=&  {1 \over \ell} \, \lbrack \d f , x \rbrack        
        = {1 \over \ell} \left( \d (f \, x - x \, f ) - \lbrack f , \d x \rbrack \right)     
        =  {1 \over \ell} ( \d x \, f - f \, \d x )        \nonumber \\        
       &=& {1 \over \ell} \lbrack f(x + \ell) - f(x) \rbrack  \, \d x 
                              \label{df_dx_1dlatt}
\eea
so that the {\em left partial derivative} defined via the first equality turns 
out to be the right discrete derivative, i.e.,
\be
    \partial_{+x} f = {1 \over \ell} \lbrack f(x + \ell) - f(x) \rbrack  \; .
\ee
Introducing a {\em right partial derivative} via $\d f = \d x \, \partial_{-x} 
f$, an application of (\ref{dx_f_1dlatt}) shows that it is the left discrete 
derivative, i.e.,
\be
    \partial_{-x} f = {1 \over \ell} \lbrack f(x) - f(x - \ell) \rbrack  \; .
\ee
An {\em indefinite integral} should have the property
\be
    \int \d f = f + \mbox{`constant'}
\ee
where `constants' are functions annihilated by $\d$. These are just the functions
with period $\ell$ (so that $f(x+\ell)=f(x)$). It turns out that every function
can be integrated.\cite{DMH92_lq} 
Since the indefinite integral is only determined up to the addition of an arbitrary
function with period $\ell$, it defines a {\em definite integral} only if the region
of integration is an interval the length of which is a multiple of $\ell$ (or a 
union of such intervals). Then one obtains
\be
     \int_{x_0 - m \ell}^{x_0 + n \ell} f(x) \, \d x 
   = \ell \, \sum_{k=-m}^{n-1} f(x_0 + k \ell)      
\ee  
and in particular 
\be
     \int_{x_0 - \infty \ell}^{x_0 + \infty \ell} f(x) \, \d x 
  =  \ell \, \sum_{k=-\infty}^{\infty} f(x_0 + k \ell)   \; .     \label{infint_1dlatt}
\ee
The integral simply picks out the values of $f$ on a {\em lattice} with 
spacings $\ell$ and forms the Riemann integral for the corresponding piecewise 
constant function on $\mathbb{R}$. The point $x_0 \in \mathbb{R}$ determines
how the lattice is embedded in $\mathbb{R}$.
                                 
\vskip.3cm

Let now $\M = {\mathbb{Z}}^n$. For $a,b \in \M$ we define a differential calculus
by
\be
                e_{ab} \neq 0 \quad \Longleftrightarrow \quad 
                b = a + \hat{\mu}     \qquad \mbox{where} \quad
               \hat{\mu} = \left( \delta^\nu_\mu \right)     \; .
\ee
This corresponds to the oriented lattice graph drawn in Fig. 3.

\begin{minipage}{4cm}
\unitlength1.cm
\begin{picture}(4,6.5)(0.5,-0.7)
\thicklines
%\linethickness{0.3mm}
%
\multiput(0,0)(1,0){6}{\circle*{0.15}}
\multiput(0,1)(1,0){6}{\circle*{0.15}}
\multiput(0,2)(1,0){6}{\circle*{0.15}}
\multiput(0,3)(1,0){6}{\circle*{0.15}}
\multiput(0,4)(1,0){6}{\circle*{0.15}}
\multiput(0,5)(1,0){6}{\circle*{0.15}}
\multiput(0,0)(1,0){5}{\vector(1,0){0.9}}
\multiput(0,1)(1,0){5}{\vector(1,0){0.9}}
\multiput(0,2)(1,0){5}{\vector(1,0){0.9}}
\multiput(0,3)(1,0){5}{\vector(1,0){0.9}}
\multiput(0,4)(1,0){5}{\vector(1,0){0.9}}
\multiput(0,5)(1,0){5}{\vector(1,0){0.9}}
\multiput(0,0)(0,1){5}{\vector(0,1){0.9}}
\multiput(1,0)(0,1){5}{\vector(0,1){0.9}}
\multiput(2,0)(0,1){5}{\vector(0,1){0.9}}
\multiput(3,0)(0,1){5}{\vector(0,1){0.9}}
\multiput(4,0)(0,1){5}{\vector(0,1){0.9}}
\multiput(5,0)(0,1){5}{\vector(0,1){0.9}}
\end{picture}
\end{minipage}
\hfill
\centerline{
\begin{minipage}[t]{2.3cm}
\centerline{\bf Fig. 3}
\vskip.1cm \noindent
\small
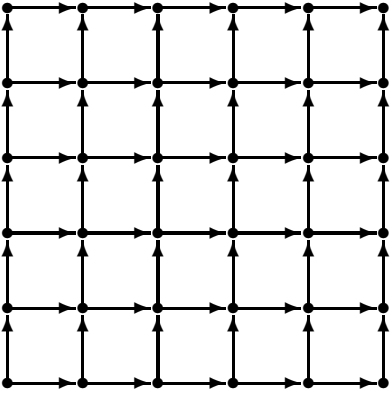
A finite part of the oriented lattice graph. 
\end{minipage}    }
 
\noindent
Introducing the lattice coordinate functions
\be
          x^\mu := \ell^\mu \, \sum_a a^\mu \, e_a
\ee
with constants $\ell^\mu > 0$, one obtains
\be                  \label{n-dim_latt_calc}
    \lbrack \d x^\mu , x^\nu \rbrack 
  = \ell^\mu \, \delta^{\mu \nu} \, \d x^\mu   
\ee
using (\ref{e_e_delta_e}). $\{ \d x^\mu \}$ is a left $\A$-module
basis of $\O^1$. The above commutation relations define
a deformation of the ordinary differential calculus on  ${\mathbb{R}}^n$.
This deformed differential calculus may be regarded as a basic structure
underlying lattice field theories.\cite{DMHS93}

\section{A class of noncommutative differential calculi on a commutative
algebra}
Let $\A$ be the associative and commutative algebra over $\mathbb{R}$ freely 
generated by elements $x^\mu$, $\mu = 1, \ldots, n$.  
  For example, the $x^\mu$ could be the canonical coordinates on ${\mathbb{R}}^n$.
The ordinary differential calculus on $\A$ has the property 
$\lbrack \d x^\mu , x^\nu \rbrack = 0$, i.e., differentials and functions
commute. Relaxing this property, there is a class of {\em noncommutative 
differential calculi} such that\footnote{On the rhs of this equation we are
using the summation convention.}
\be                 \label{dx_x_C}
 \lbrack \d x^\mu , x^\nu \rbrack = C^{\mu \nu}{}_\kappa \, \d x^\kappa
\ee
with {\em structure functions} $C^{\mu \nu}{}_\kappa(x^\lambda)$ which
have to satisfy some consistency conditions. First, we have
\bea
      \lbrack \d x^\mu , x^\nu \rbrack 
  &=& (\d x^\mu) \, x^\nu - x^\nu \, \d x^\mu  \nonumber \\
  &=& \d ( x^\mu x^\nu - x^\nu x^\mu ) 
      - x^\mu \, \d x^\nu + (\d x^\nu) \, x^\mu    
   =  \lbrack \d x^\nu , x^\mu \rbrack   \; .
\eea
Assuming the differentials $\d x^\mu$, $\mu = 1, \ldots, n$, to be linearly
independent\footnote{More precisely, we assume here that the $\d x^\mu$ form 
a left $\A$-module basis of $\Omega^1(\A)$.}, this implies
\be
       C^{\mu \nu}{}_\kappa = C^{\nu \mu}{}_\kappa  \; .     \label{consist_cond1}
\ee  
  Furthermore,
\bea
 0 &=& \left( \lbrack \d x^\mu , x^\nu \rbrack 
       - C^{\mu \nu}{}_\kappa \, \d x^\kappa \right) \, x^\lambda
       =  \lbrack (\d x^\mu) \, x^\lambda , x^\nu \rbrack - C^{\mu \nu}{}_\kappa 
           \, (\d x^\kappa) \, x^\lambda   \nonumber \\
   &=& \lbrack x^\lambda \, \d x^\mu + C^{\mu \lambda}{}_\rho \, \d x^\rho, 
       x^\nu \rbrack - C^{\mu \nu}{}_\kappa \, ( x^\lambda \, \d x^\kappa 
       + C^{\kappa \lambda}{}_\rho \, \d x^\rho ) \nonumber \\
   &=& x^\lambda \, \lbrack \d x^\mu , x^\nu \rbrack + C^{\mu \lambda}{}_\rho 
       \, \lbrack \d x^\rho , x^\nu \rbrack - x^\lambda \, C^{\mu \nu}{}_\kappa 
       \, \d x^\kappa - C^{\mu \nu}{}_\kappa \, C^{\kappa \lambda}{}_\rho \, 
       \d x^\rho    \nonumber \\
   &=& ( C^{\mu \lambda}{}_\rho \, C^{\rho \nu}{}_\kappa - C^{\mu \nu}{}_\rho 
       \, C^{\rho \lambda}{}_\kappa ) \, \d x^\kappa 
\eea
which leads to
\be                     \label{consist_cond2}
   C^{\lambda \mu}{}_\rho \, C^{\nu \rho}{}_\kappa = C^{\nu \mu}{}_\rho 
       \, C^{\lambda \rho}{}_\kappa 
\ee
or, in terms of the matrices $C^\mu$ with entries $(C^\mu)^\nu{}_\kappa :=
C^{\mu \nu}{}_\kappa$,
\be                     \label{consist_cond2'}
           C^\mu \, C^\nu = C^\nu \, C^\mu  \; .
\ee
  For constant $C^{\mu \nu}{}_\kappa$ and $n \leq 3$, a classification of all 
solutions of the consistency conditions (\ref{consist_cond1}) and 
(\ref{consist_cond2}) has been obtained.\cite{DMHS93,BDMH95}
Besides the `lattice differential calculus' discussed in the previous subsection, 
this includes other interesting deformations of the ordinary differential 
calculus on ${\mathbb{R}}^n$.\cite{DMH93_tcalc,DMH93_stoch}
The relations (\ref{n-dim_latt_calc}) are obviously not invariant under (suitable)
coordinate transformations. Invariance is achieved, however, with the
form (\ref{dx_x_C}) of the commutation relations.
\vskip.3cm

  From the structure functions we can build
\be
       g^{\mu \nu} := \mbox{Tr} ( C^\mu \, C^\nu )   \label{C-metric}
\ee
which for the lattice calculus (\ref{n-dim_latt_calc}) becomes 
$(\ell^\mu)^2 \, \delta^{\mu \nu}$, a kind of metric tensor. In the framework
under consideration, the metric arises as a composed object. The set of structure
functions $C^{\mu \nu}{}_\kappa$ is the more fundamental geometric
structure.

\section{An application in the context of integrable models}
  For  two-dimensional $\sigma$-models there is a construction of an infinite sequence
of conserved currents \cite{BIZZ79} which can be formulated very compactly
in terms of ordinary differential forms. This then suggests to generalize the notion 
of a $\sigma$-model to noncommutative differential calculi such that the 
construction of conservation laws still works. In this way one obtains a simple though 
very much non-trivial application of the formalism of noncommutative 
geometry.\cite{DMH96_intd,DMH97_intm}

\subsection{Generalized integrable $\sigma$-models}
Let $\cal A$ be an associative and commutative\footnote{A generalization of the 
following to {\em non}commutative algebras seems to be possible if they admit an 
involution ${}^\dag$. Then (\ref{star-covariance}) has to be replaced by 
$ \star \, (w \, f) = f^\dag \, \star \, w$.}
algebra with unit $\idty$ and $(\Omega,\d )$ a differential calculus on it. 
 Furthermore, let $\star \, : \; \Omega^1 \rightarrow \Omega^1$ be an 
invertible linear map such that
\be              \label{star-covariance}
         \star \, (w \, f) = f \, \star \, w 
\ee
and
\be              \label{star-symm}
                  w \star w' = w' \star w   \; .
\ee
In addition, we require that
\be              \label{closed-star-star} 
   \d w = 0 \quad \Rightarrow \quad w = \star \star \, \d \chi
\ee
with $\chi \in {\cal A}$. 
 Furthermore, let $a \in GL(n,{\cal A})$ and $A := a^{-1} \, \d a$. Then
\be
     F := \d A + A A \equiv 0                      \label{F=0}
\ee
since $\d a^{-1} = - a^{-1} \, (\d a) \, a^{-1}$. These definitions are
made in such a way that the field equation of a {\em generalized 
$\sigma$-model}
\be
     \d \star A = 0      \label{field_eq}
\ee
and a construction of an infinite set of conservation laws in two 
dimensions\cite{BIZZ79} generalizes to a considerably more general 
framework. Let $\chi$ be an $n \times n$ matrix  with entries in $\cal A$.
Using the two relations (\ref{star-covariance}), (\ref{star-symm}), 
and  the field equation $\d \star A = 0$, we find
\bea
  \d \star (A^i{}_j \, \chi^j{}_k) = \d (\chi^j{}_k \, \star A^i{}_j)
  = (\d \chi^j{}_k) \, \star A^i{}_j + \chi^j{}_k \, \d \star A^i{}_j
  = A^i{}_j \star \d \chi^j{}_k    \quad
\eea
and thus 
\bea
\d \star D \chi = \d \star \d \chi + \d (\star A \chi) 
 = \d \star \d \chi + A \star \d \chi = D \star \d \chi    \label{Lemma} 
\eea
where $D \chi := \d \chi + A \, \chi$. Let 
\bea
    \chi^{(0)} := \left( \begin{array}{cccc}
                  \idty & 0 & \cdots & 0           \\
                   0    & \ddots & \ddots & \vdots \\
                 \vdots & \ddots & \ddots & 0      \\
                   0    & \cdots &   0    & \idty
                  \end{array} \right)   
\eea
Then 
\be
    J^{(1)} := D \chi^{(0)} = A
\ee
so that
\be
    \d \star \, J^{(1)} = 0
\ee
as a consequence of the field equation. Hence, using (\ref{closed-star-star}),
\be
     J^{(1)} = \star \, \d \chi^{(1)}
\ee 
with a matrix $\chi^{(1)}$. Now, let $J^{(m)}$ be a conserved current, i.e.,
\be
     J^{(m)} = \star \, \d \chi^{(m)}  \; .     \label{J-first}
\ee
Then
\be
     J^{(m+1)} := D \chi^{(m)}  \qquad   (m \geq 0)  \label{J-second}
\ee
is also conserved since
\bea
     \d \star J^{(m+1)} &=& \d \star D \chi^{(m)} = D \star \d \chi^{(m)} 
   = D J^{(m)} = D^2 \chi^{(m-1)} \nonumber \\
   &=& F \, \chi^{(m-1)} = 0      \qquad (m \geq 1) 
\eea
using (\ref{Lemma}) and the identity (\ref{F=0}).
Starting with $J^{(1)}$, we obtain an infinite set of (matrices of) conserved 
currents.\footnote{There is no guarantee, however, that all these currents are really 
independent. For example, our formalism includes the free linear wave equation on 
two-dimensional Minkowski space. In that case, the higher conserved charges are 
just polynomials in the first one (which is the total momentum). }
In case of the ordinary differential calculus on a two-dimensional Riemannian
manifold, this construction reduces to the classical one.\cite{BIZZ79}
\vskip.3cm

Let us (formally) define
\be
     \chi := \sum_{m \geq 0} \lambda^m \, \chi^{(m)}  
\ee 
with a constant $\lambda \neq 0$. Then (\ref{J-first}) and (\ref{J-second}) 
lead to
\be
     \star \, \d \chi = \lambda \, D \chi   \; .   \label{lin_eq}
\ee
As a consequence of this equation we get
\be
     0 = \d \star D \chi^i{}_j = D \star \d \chi^i{}_j 
         + \chi^k{}_j \, \d \star A^i{}_k    
\ee
and
\be
  D \star \d \chi = \lambda \, D^2 \chi = \lambda \, F \, \chi  \; .    
\ee
Using $A=a^{-1} \, \d a$, the integrability condition of the linear equation
(\ref{lin_eq}) is the field equation (\ref{field_eq}).
\vskip.3cm

We have extended the definition of a class of $\sigma$-models
to a rather general framework of noncommutative geometry, though still with the
restriction to a commutative algebra $\A$ (which can always be realized as an algebra
of functions on some space), but with {\em non}commutative differential calculi 
(where functions and differentials do not commute, in general). Already in this
case a huge set of possibilities for integrable models 
appears.\cite{DMH96_intd,DMH97_intm}

\subsection{Example: recovering the Toda lattice}
Let $\cal A$ be the (commutative) algebra of functions on 
$\M = \ell_0 {\mathbb{Z}} \times \ell_1 \mathbb{Z}$. Here $\ell_k \mathbb{Z}$
stands for the one-dimensional lattice with spacings $\ell_k > 0$. A special
differential calulus $(\O(\A), \d )$ on $\cal A$ is then determined by the 
following commutation relations,
\be             \label{comm_rels}
   \lbrack \d t , t \rbrack = \ell_0 \, \d t   \, , \quad 
   \lbrack \d x , x \rbrack = \ell_1 \, \d x  \, , \quad
   \lbrack \d t , x \rbrack = \lbrack \d x , t \rbrack = 0 
\ee
where $t$ and $x$ are the canonical coordinate functions on $\ell_0 \mathbb{Z}$ and
$\ell_1 \mathbb{Z}$, respectively. This is our lattice differential calculus 
(\ref{n-dim_latt_calc}) for $n=2$. As a consequence, we have
\be
   \d t \, f({\bf x}) = f({\bf x} + \ell_0) \, \d t \, , \quad
   \d x \, f({\bf x}) = f({\bf x}+\ell_1) \, \d x
\ee
where 
\be
     {\bf x} := (t,x) \, , \quad 
     {\bf x}+\ell_0 := (t+\ell_0,x) \, , \quad 
     {\bf x}+\ell_1 := (t,x+\ell_1)  \; .
\ee
 Furthermore,
\be
   \d f = {1 \over \ell_0} \lbrace f({\bf x}+\ell_0) -f({\bf x}) \rbrace \, \d t 
            + {1 \over \ell_1} \lbrace f({\bf x}+\ell_1) -f({\bf x}) \rbrace \,  \d x  \; .
\ee
Acting with $\d$ on (\ref{comm_rels}), we obtain
\be
    \d t \, \d x = - \d x \, \d t \, , \quad 
    \d t \, \d t = 0 = \d x \, \d x  \; .
\ee
This familiar anticommutativity of differentials does not extend 
to general 1-forms, however. The differential calculus has the following
property.

\vskip.2cm
\noindent
{\em Lemma.} Every closed 1-form is exact. 
\vskip.1cm
\noindent
{\em Proof:} 
  For $w=w_0(t,x) \, \d t + w_1(t,x) \, \d x$ the condition $\d w=0$ means
$\pa_{+t}w_1 = \pa_{+x}w_0$. For simplicity, we set $\ell_0 = \ell_1 =1$
in the following. Let us define\footnote{This function is obtained by integrating
$w$ along a path $\gamma \, : \, {\mathbb{N}} \rightarrow {\mathbb{Z}}^2$ first
from $(0,0)$ to $(t,0)$ along the $t$-lattice direction, then from $(t,0)$ to $(t,x)$
along the $x$-lattice direction. The result does not dependent on the chosen path.
This follows from an application of Stokes' theorem.}
$$ 
   F(t,x) := \sum_{k=0}^{t-1} w_0(k,0) + \sum_{j=0}^{x-1} w_1(t,j) \; .
$$
It satisfies
$$ 
 \pa_{+x}F = F(t,x+1)-F(t,x) 
                   = \sum_{j=0}^x w_1(t,j) - \sum_{j=0}^{x-1} w_1(t,j)
                   = w_1(t,x)
$$ 
and, using $\d w=0$, 
\begin{eqnarray*}
 \pa_{+t}F &=& F(t+1,x)-F(t,x) = w_0(t,0) + \sum_{j=0}^{x-1}\pa_{+t}w_1(t,j)   \\
 &=& w_0(t,0) + \sum_{j=0}^{x-1}\pa_{+x}w_0(t,j)
 = w_0(t,0) + w_0(t,x) - w_0(t,0) = w_0(t,x)  \; .
\end{eqnarray*}
Hence $w = \d F$.
                                        \hfill  $\blacksquare$
\vskip.2cm

Let us now turn to the conditions for the $\star$-operator. First we introduce
$g^{\mu \nu}$ via
\be
      \d x^\mu \star \d x^\nu = g^{\mu \nu} \, \d t \, \d x  \; .
\ee 
With $w = w_\mu \, \d x^\mu$,  (\ref{star-symm}) becomes
\be
   \lbrack w_\mu({\bf x}) \, w'_\nu({\bf x}+\ell_\mu-\ell_\nu) -  w'_\mu({\bf x}) \, 
   w_\nu({\bf x}+\ell_\mu-\ell_\nu) \rbrack \, g^{\mu \nu} = 0 
 \qquad \forall \, w_\mu, w'_\mu  \; .
\ee
It yields
\be
       g^{\mu \nu} = c^\mu \, \delta^{\mu \nu}
\ee
with arbitrary functions $c^\mu({\bf x})$ which have to be different from zero in order
for $\star$ to be invertible. This includes the metric (\ref{C-metric}). For the generalized 
Hodge operator we now obtain
\be
    \star \, \d t = c^0({\bf x}-\ell_0) \, \d x \, , \qquad 
    \star \, \d x = - c^1({\bf x}-\ell_1) \, \d t   
\ee
which extends to $\Omega^1$ via (\ref{star-covariance}).
\vskip.1cm

In the following, we choose $g^{\mu \nu} = \eta^{\mu \nu}$ which in classical differential
geometry are the components of the two-dimensional Minkowski metric with respect to 
an inertial coordinate system. 
We then have $\star \star w({\bf x}) = w({\bf x}-\ell_0 - \ell_1)$ which, together with the 
above Lemma, implies (\ref{closed-star-star}). 
Therefore, the construction of conservation laws does work in the case 
under consideration. Let us look at the simplest generalized $\sigma$-model where 
$a$ is just a function (i.e., a $1 \times 1$-matrix). We write
\be
      a = e^{- q(t,x)}
\ee
with a function $q$ and $q_k(n) := q(n \ell_0, k \ell_1)$. Then
\bea 
      A &=& {1 \over \ell_0} (e^{q_k(n)-q_k(n+1)}-1) \, \d t
             + {1 \over \ell_1} (e^{q_k(n)-q_{k+1}(n)}-1) \, \d x       \\
     \ast A &=& -{1 \over \ell_0} (e^{q_k(n-1)-q_k(n)}-1) \, \d x
                  -{1 \over \ell_1} (e^{q_{k-1}(n)-q_k(n)}-1) \, \d t 
\eea
and the field equation $\d \star A = 0$ takes the form
\bea
    {1 \over \ell_0^2} \left [ e^{q_k(n-1)-q_k(n)}-e^{q_k(n)- 
     q_k(n+1)} \right ]
   = {1 \over \ell_1^2}  \left[ e^{q_{k-1}(n)-q_k(n)}
     - e^{q_k(n)-q_{k+1}(n)} \right ]    \quad 
\eea
Replacing $\A$ with the algebra of functions on ${\mathbb{R}} \times \ell_1 \mathbb{Z}$
which are smooth in the first argument, the limit $\ell_0 \to 0$ can be performed.
This contraction leads to
\be                  \label{nonlin_Toda}
   \ddot{q}_k + {1 \over \ell_1^2} (e^{q_k-q_{k+1}} - e^{q_{k-1} - q_k}) = 0
\ee 
which is the {\em nonlinear Toda lattice} equation \cite{Toda89}. 
In particular, in this way a new and simple understanding of its complete integrability 
has been achieved. There is a  `noncommutative geometry' naturally associated with
the Toda lattice equation.
Generalizations of the Toda lattice are obtained by replacing the function 
$a$ with a $GL(n,{\cal A})$-matrix.\cite{DMH96_intd}

\section{Metrics in noncommutative geometry}
In the previous section we have introduced a generalized Hodge 
$\star$-operator. In classical (Riemannian) differential geometry, the 
Hodge operator contains information equivalent to a  metric tensor
which in turn has its origin in the problem of defining the length of a 
curve and the distance between points of a Riemannian space. 
On the basis of the formalism sketched in section 3.1, Connes proposed 
a generalization of the classical distance function
to discrete and even noncommutative spaces.\cite{Conn94} 
Some examples are discussed in the following subsection.\cite{DMH97_dist} 
The relation with a generalized Hodge operator or other generalized concepts 
of a metric still has to be understood, however.

\subsection{Connes' distance function associated with differential calculi on 
 finite sets}
Let $(\A, {\cal H}, \hat{\cal D})$ be a spectral triple (cf section 3.1). A {\em state} 
on $\A$ is a linear map $\phi \, :\, \A \rightarrow \mathbb{C}$ which is positiv, i.e., 
$\phi( a^\ast a ) \geq 0 $ for all $a \in \A$, and normalized, i.e., $\phi(\idty) = 1$. 
According to Connes \cite{Conn94}, the distance between two states $\phi$ and 
$\phi'$ is given by
\be
     d(\phi,\phi') := \mbox{sup} \lbrace | \phi(a) - \phi'(a) | \; ; \,
                      a \in \A, \, \| \lbrack \hat{\cal D} , a \rbrack \|
                      \leq 1 \rbrace   \; .
\ee 
Given a set $\cal M$, each point $p \in \cal M$ defines a state $\phi_p$
via $\phi_p(f) := f(p)$ for all functions $f$ on $\cal M$. The above formula
then becomes
\be
     d(p,p') := \mbox{sup} \lbrace | f(p) - f(p') | \; ; \,
                      f \in \A, \, \| \lbrack \hat{\cal D} , f \rbrack \|
                      \leq 1 \rbrace   \; .
\ee 
\vskip.3cm
\noindent
{\em Example 1.} The universal first order differential calculus on a set
of two elements $p,q$ is described by a graph consisting of two points 
which are connected by a pair of antiparallel arrows. Its adjacency 
matrix is
\be
            {\cal D} = \left( \begin{array}{cc} 0 & 1 \\
                                                1 & 0
                              \end{array} \right)  
\ee
so that
\be
    \lbrack {\cal D} , f \rbrack =
    \left( \begin{array}{cc} 0 & f(p)-f(q) \\ 
                            f(q)-f(p) & 0
                              \end{array} \right)    \; .
\ee
Then
\bea
         \| \lbrack {\cal D} , f \rbrack \|^2 
     &=& \mbox{sup}_{\| \psi \| = 1} 
         \| \lbrack {\cal D} , f \rbrack \, \psi \|^2 
     = \mbox{sup}_{\| \psi \| = 1} | f(p)-f(q) |^2 \, 
        ( |\psi_1|^2 + |\psi_2|^2 )  \nonumber \\
     &=& | f(p)-f(q) |^2
\eea
for $\psi \in {\mathbb{C}}^2$.
It follows that Connes' distance function defined with the adjacency
matrix gives $d(p,q)=1$. In this example, which appeared 
in models of elementary particle physics \cite{Conn+Lott91}, 
there is no need for a `doubling' of the representation as in (\ref{doubled_D,f}). 
We may, however, replace $\cal D$ by $\hat{\cal D}$ also in this case.
The result for the distance between the two points remains unchanged,
however.    

\vskip.3cm
\noindent
{\em Example 2.} \cite{DMH97_dist} Let us consider the first order 
differential calculus on a set of $N$ elements determined by the 
graph in Fig. 4. 

\begin{minipage}{4cm}
\unitlength1.cm
\begin{picture}(4,1)(0.5,0)
\thicklines
%\linethickness{0.3mm}
%
\multiput(0,0)(1,0){6}{\circle*{0.15}}
\multiput(0,0)(1,0){3}{\vector(1,0){0.9}}
\put(3.3,0){$\ldots$}
\multiput(4,0)(1,0){1}{\vector(1,0){0.9}}
\end{picture}
\end{minipage}
\hfill
\centerline{
\begin{minipage}[t]{3cm}
\centerline{\bf Fig. 4}
\vskip.1cm \noindent
\small
A finite oriented linear lattice graph. 
\end{minipage}    }

\vskip.3cm
\noindent
The corresponding adjacency matrix is
\bea
       \left( \begin{array}{cccr} 0 & 1 & 0 & \cdots  0 \\
                                       \vdots & \ddots & \ddots & \vdots \\
                                       \vdots &  & \ddots & 1 \\
                                              0 & \cdots & \cdots  & 0
                \end{array} \right)   \; .
\eea
This matrix contains all the topological information about the lattice, i.e., the
neighbourhood relationships. We can add information about the distances between
neighbouring points to it in the following way. Let $\ell_k$ be the distance from
point $k$ to point $k+1$  (numbering  the lattice sites by $1, \ldots, N$). We define
\bea
     {\cal D}_N := \left( \begin{array}{cccr} 0 & \ell_1^{-1} & 0 & \cdots  0 \\
                                                             \vdots & \ddots & \ddots & \vdots \\
                                                             \vdots &  & \ddots & \ell_{N-1}^{-1} \\
                                                                    0 & \cdots & \cdots  & 0
                                    \end{array} \right)   \; .
\eea
With a complex function $f$ we associate a real function $F$ via
\be
     F_1 := 0 \, , \quad F_{i+1} := F_i + | f_{i+1} - f_i | \qquad 
                i = 1, \ldots, N-1  \; .
\ee
where $f_i := f(i)$. Then $ | F_{i+1} - F_i | = | f_{i+1} - f_i | $ and
\be
      \| \lbrack \hat{\cal D}_N , \hat{f} \rbrack \, \psi \|
  = \| \lbrack \hat{\cal D}_N , \hat{F} \rbrack \, \psi \| 
\ee
for all $\psi \in {\mathbb{C}}^{2N}$.
Hence, in calculating the supremum over all functions $f$ in
the definition of Connes' distance function, it is sufficient to consider
{\em real} functions. Then $Q_N := \lbrack \hat{\cal D}_N , \hat{f} \rbrack$ 
is anti-hermitean and its norm is then given by the maximal absolute value
of its eigenvalues. Instead of $Q_N$ it is simpler to consider
$Q_N \, Q_N^\dagger$ which is already diagonal with entries
$0, \ell_1^{-2} (f_2-f_1)^2, \ldots, \ell_{N-1}^{-2} (f_N-f_{N-1})^2, 
\ell_1^{-2} (f_2-f_1)^2, \ldots, \ell_{N-1}^{-2} (f_N-f_{N-1})^2, 0$ 
on the diagonal. This implies
\be
   \| \lbrack \hat{\cal D}_N , \hat{f} \rbrack \| 
 = \mbox{max} \, \lbrace \ell_1^{-1}| f_2-f_1 | , \ldots, \ell_{N-1}^{-1} 
    | f_N-f_{N-1} | \rbrace   \; .
\ee
We have the obvious inequality
\be
   d(i,i+k) \leq \mbox{sup} \lbrace | f(i+1) - f(i) | + \ldots + | f(i+k) - f(i+k-1) | \, ;
   \,  \| \lbrack \hat{\cal D}_N , \hat{f} \rbrack \| \leq 1 \rbrace  \; .
\ee
But a closer inspection shows that actually equality holds here. We conclude that 
$ d(i,i+k) = \ell_i + \ell_{i+1} + \ldots + \ell_{i+k-1}$. \footnote{A different choice
of $\hat{\cal D}$ has been made elsewhere \cite{BLS94} to define the 
distance on a lattice. See also Rieffel \cite{Rief93} for a reformulation of discrete 
metric spaces in Connes' framework.}
                                                                                      \hfill  $\blacksquare$

\vskip.3cm
\noindent
{\em Example 3.} The graph in Fig. 1 has the adjacency matrix
\bea
     {\cal D} =  \left( \begin{array}{cccc} 0 & 1 & 0 & 1 \\
                                                               0 & 0 & 1 & 0 \\
                                                               0 & 0 & 0 & 0 \\
                                                               0 & 0 & 1 & 0
                       \end{array} \right)   \; .
\eea
The norm of $\lbrack \hat{\cal D} , \hat{f} \rbrack$ is the positive 
square root of the largest eigenvalue of  
\bea
   \lbrack \hat{\cal D} , \hat{f} \rbrack \, \lbrack \hat{\cal D} , \hat{f} \rbrack^\dagger 
   = \left( \begin{array}{cc}
       \lbrack {\cal D} , f^\ast \rbrack^\dagger \,  \lbrack {\cal D} , f^\ast \rbrack &  0 \\
      0 & \lbrack {\cal D} , f \rbrack \,  \lbrack {\cal D} , f \rbrack^\dagger
                       \end{array} \right)   \; .
\eea
It follows that
\be
  \| \lbrack \hat{\cal D} , \hat{f} \rbrack \| = \mbox{max} \lbrace 
  \sqrt{ | f_{21}|^2 + | f_{41}|^2},  \sqrt{ | f_{32}|^2 + | f_{34}|^2} \rbrace 
\ee
where $f_{kl} := f_k - f_l $. Introducing 
\be
     x := {1 \over 2} ( f_{21}-f_{41}) \, , \quad
     y := {1 \over 2} ( f_{21}+f_{41}) \, , \quad
     z := {1 \over 2} ( f_{32}+f_{34}) \, , 
\ee
we find the `Euclidean' result
\be
    d(1,3) = \mbox{sup} \lbrace {1 \over 2} |y+z| \, ; \, \mbox{max} \lbrace |x|^2+|y|^2 ,
                 |x|^2+|z|^2 \rbrace \leq 2 \rbrace 
              = \sqrt{2}  \; . 
\ee
                                     \hfill  $\blacksquare$

\vskip.3cm
\noindent
{\em Example 4.} Let us consider the following digraph which is part of
the lattice graph in Fig. 3.
 
\begin{minipage}{7cm}
\begin{center}
\unitlength=1.5cm     
\begin{picture}(4.,1.8)(-0.5,-0.4)
\thicklines
\multiput(0,0)(1,0){3}{\circle*{0.1}}
\multiput(0,1)(1,0){3}{\circle*{0.1}}
\put(-0.2,-0.25) {$1$}
\put(0.95,-0.25) {$2$}
\put(2.15,-0.25) {$3$}
\put(2.15,1.1) {$4$}
\put(0.95,1.1) {$5$}
\put(-0.2,1.1) {$6$}
\multiput(0,0)(1,0){2}{\vector(1,0){0.9}}
\multiput(0,0)(1,0){3}{\vector(0,1){0.9}}
\multiput(0,1)(1,0){2}{\vector(1,0){0.9}}
\end{picture}
\end{center}
\end{minipage}
\hfill
\begin{minipage}{4cm}
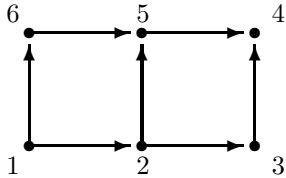
\centerline{\bf Fig. 5}
\small
\noindent
The digraph associated with a special first order differential calculus on a set 
of six points.
\end{minipage}

\noindent
A numerical evaluation of the distance function shows that $d(3,6) \lesssim 2 < d(1,4) 
< \sqrt{5}$ and thus deviates from the Euclidean value.
                                                                            \hfill  $\blacksquare$
\vskip.2cm

The last example shows that Connes' distance function defined in terms of the adjacency 
matrix of the $n$-dimensional oriented lattice graph with $n > 1$ (see Fig. 3) does 
{\em not} assign to it a Euclidean geometry, as might have been conjectured on the basis 
of our examples 2 and 3.

\section{Final remarks}
This work centered around examples which live on lattices. Such
spaces do not at all exhaust the possibilities of noncommutative geometry
of commutative algebras. In this case, and more generally in the case
of discrete spaces, the generalized partial derivatives of a differential
calculus are {\em discrete} derivatives, corresponding to an infinite sum
of powers of ordinary partial derivatives. There are other differential
calculi where the generalized partial derivatives are differential operators
of finite order and some of them appear to be of relevance for an 
analysis of soliton equations, for example.\cite{DMH96_sol}
There is much more to mention in this context and we refer to a recent 
review \cite{MH97} for further information and a guide 
to the relevant literature. 
\vskip.3cm

An important aspect of the formalism of noncommutative geometry is a technical
one. On the level of generalized differential forms we have very compact expressions
which are easy to handle thanks to the simple rules of differential calculus. 
Decomposed into components, however, we end up with rather complicated
formulas, in general. This is precisely the experience which especially relativists make 
when they encounter the Cartan formalism in general relativity. Our generalization
of the construction of conserved currents for (generalized) $\sigma$-models 
reviewed in section 6.1 is another nice example.

\section*{Acknowledgments} 
  F M-H would like to thank the organizers of the 21st Johns Hopkins Workshop
and in particular Professor Yishi Duan for the kind invitation and an enjoyable 
time in Lanzhou.

\end{document}